\documentclass[aps,prd,
twocolumn,
superscriptaddress,
nofootinbib
]{revtex4-1}
\usepackage{graphicx}
\usepackage{amsmath}
\usepackage{amssymb}
\usepackage{amsthm}
\usepackage[usenames,dvipsnames]{color}
\newcommand{\sout}[1]{}
\usepackage{soul}

\newcommand{\norm}[1]{\|#1\|}
\newcommand{\Q}{\norm{\delta h}/\norm{h}}
\newcommand{\NWIP}[2]{ \langle{#1},{#2}\rangle }    

\makeatletter
\newcommand{\etal}{\textit{et~al}\@ifnextchar{\relax}{.\relax}{\ifx\@let@token.\else\ifx\@let@token~.\else.\@\xspace\fi\fi}}
\makeatother

\newcommand{\CITA}{\affiliation{Canadian Institute for Theoretical
    Astrophysics, University of Toronto, 60~St.~George Street,
    Toronto, Ontario M5S 3H8, Canada}} %
\newcommand{\Caltech}{\affiliation{Theoretical Astrophysics 350-17,
    California Institute of Technology, Pasadena, CA 91125, USA}} %
\newcommand{\Cornell}{\affiliation{Center for Radiophysics and Space
    Research, Cornell University, Ithaca, New York, 14853, USA}}

\begin{document}


\title{Suitability of hybrid gravitational waveforms for unequal-mass binaries}

\author{Ilana MacDonald}
\email[]{macdonald@astro.utoronto.ca}

\author{Abdul H. Mrou\'e}

\author{Harald P. Pfeiffer}
\CITA

\author{Michael Boyle}
\Cornell

\author{Lawrence E. Kidder}
\Cornell

\author{Mark A. Scheel}

\author{B\'ela Szil\'agyi}

\author{Nicholas W. Taylor}
\Caltech

\date{\today}

\begin{abstract}
  This article studies sufficient accuracy criteria of hybrid
  post-Newtonian (PN) and numerical relativity (NR) waveforms for
  parameter estimation of strong binary black-hole sources in
  second-generation ground-based gravitational-wave detectors. We
  investigate equal-mass non-spinning binaries with a new 33-orbit NR
  waveform, as well as unequal-mass binaries with mass ratios $2,3,4$
  and $6$.  For equal masses, the 33-orbit NR waveform allows us to
  recover previous results and to extend the analysis toward matching
  at lower frequencies.  For unequal masses, the errors between
  different PN approximants increase with mass ratio.  Thus, at 3.5PN,
  hybrids for higher-mass-ratio systems would require NR
  waveforms with many more gravitational-wave (GW) cycles to guarantee
  no adverse impact on parameter estimation.  Furthermore, we
  investigate the potential improvement in hybrid waveforms that can
  be expected from 4th order post-Newtonian waveforms, and find that
  knowledge of this 4th post-Newtonian order would significantly
  improve the accuracy of hybrid waveforms.
\end{abstract}


\maketitle

\section{Introduction}

Within the next few years, Advanced
LIGO~\cite{Abbott:2007,Shoemaker:aLIGO,Harry2010}, 
Virgo~\cite{aVIRGO}, and LCGT~\cite{Kuroda:2010} will likely observe the
first gravitational wave signals. The detection rate
is expected to be somewhere between 0.4 and 1000 detections per
year for binary black-hole (BBH) systems with masses below
100$M_\odot$ at distances of several
gigaparsecs~\cite{Abadie:2010cfa}. Because the gravitational wave (GW) signals will be faint
compared to the detector noise, accurate source modeling of the
predicted GWs will be necessary to detect signals and perform parameter
estimation on them via matched filtering.

Several methods have been developed to calculate the GWs
from a BBH system, two of which are post-Newtonian (PN) theory and
Numerical Relativity (NR). The PN expansion~\cite{Blanchet2006} is a slow-motion, weak-field
approximation to General Relativity, and it provides an accurate
description of the inspiral prior to merger, but becomes increasingly
inaccurate close to merger. NR can be used to model the late inspiral,
merger, and ringdown of a BBH evolution, but at high computational
cost~\cite{Centrella:2010,Pfeiffer:2012pc}. 

In order to use the best features of each type of model, one stitches a
long PN inspiral to the front of an NR waveform of the late inspiral,
merger, and ringdown, thus creating a \emph{hybrid} waveform. The trick
is to stitch the NR and PN parts together such that the error due to
higher-order unknown terms in the PN waveform is small, while
necessitating as few NR orbits as possible to reduce the computational
cost. This has been explored in our previous
work~\cite{MacDonald:2011ne} for equal-mass, non-spinning binaries,
and also
in~\cite{Santamaria:2010yb,Hannam:2010,Boyle:2011dy,OhmeEtAl:2011}.

Hybrid gravitational waveforms have many uses. They play an
important role in the creation of phenomenological
waveforms~\cite{Ajith:2008b,Santamaria:2010yb}, which are used in
template banks for event detection. In addition, hybrids are used within the NINJA
project~\cite{NinjaWebPage,ninjashort,Ajith:2012tt} in order to test the GW search
pipeline for ground-based detectors.

A sufficient criterion to determine whether PN+NR hybrid waveforms are suitable
 for parameter estimation in the advanced detector era is that a GW model
 $h_\text{model}$ must be deemed \emph{indistinguishable} from the
 true waveform $h_\text{exact}$ within the Advanced LIGO noise
 spectrum. That is, we place upper bounds on the error measure $\Q$
 (to be defined in Sec.~\ref{quantifying_errors}),
 where $\delta h =  h_\text{model} - h_\text{exact}$. This error
 measure is distance-independent as long as cosmological redshifts
 are neglible.

This paper extends our earlier analysis in several directions: First,
we calculate $\Q$ for equal-mass 
hybrids constructed from a previously unpublished 33-orbit NR waveform in
order to confirm results obtained with the 15-orbit NR used in prior
works. Second, we extend the parameter space to hybrid waveforms from
unequal-mass binaries by calculating this error in the most demanding
usage case of parameter 
estimation. Third, we examine the possibility of decreasing hybrid errors if PN
were known to 4th order. A decrease in PN error would mean that
more accurate hybrid waveforms could be created with the
current length NR waveforms, thus decreasing the computational cost
of generating these templates.

This paper is organized as follows: Section~\ref{methodology} reviews
the methodology of our analysis, including the types of PN and NR
waveforms, the hybridization procedure and how the errors are
quantified. Section~\ref{extralongNR} extends our earlier results
in~\cite{MacDonald:2011ne} and~\cite{Boyle2007} using a new 33-orbit
NR waveform. Section~\ref{unequalmass} applies our error analysis to
unequal-mass binaries. Finally, section~\ref{4PNerror} estimates the
errors one would obtain for PN+NR hybrids if the 4PN terms were
known. 

\section{Methodology \label{methodology}}

\subsection{Post-Newtonian waveforms}

Post-Newtonian (PN) theory presents a slow-motion, weak-field
approximation to General Relativity in terms of expansions of
$GM/rc^2$ and $v^2/c^2$. We use the same PN approximants as in our
previous work~\cite{MacDonald:2011ne}. Specifically, we investigate
the properties of TaylorT1, TaylorT2, TaylorT3, and TaylorT4 (as
defined in~\cite{Damour2001, Buonanno-Cook-Pretorius:2007, Boyle2007})
to 3.5PN order in phase and 3.0PN order in
amplitude~\cite{Blanchet02a, BlanchetEtAl2005a, BlanchetEtAl2005b,
  Blanchet-Buonanno-Faye:2006, BlanchetEtAl:2007, Kidder:2007rt,
  Arun:2008kb, BlanchetEtAl:2010}. These four approximants differ only
in their unknown higher-order PN terms. As in~\cite{MacDonald:2011ne},
we consider the (2,2) mode of the spin-weight $s=-2$
spherical-harmonic decomposition of the gravitational waveform. The
amplitude of the PN waveform of this mode is always used at 3PN order,
except for Fig.~\ref{fig:phase_err_T3} which uses the amplitude to
2.5PN order for consistency with earlier work~\cite{Boyle2007}. 

\subsection{Numerical waveforms \label{NumericalWaveforms}}

\begin{figure}
\includegraphics[width=0.98\columnwidth]{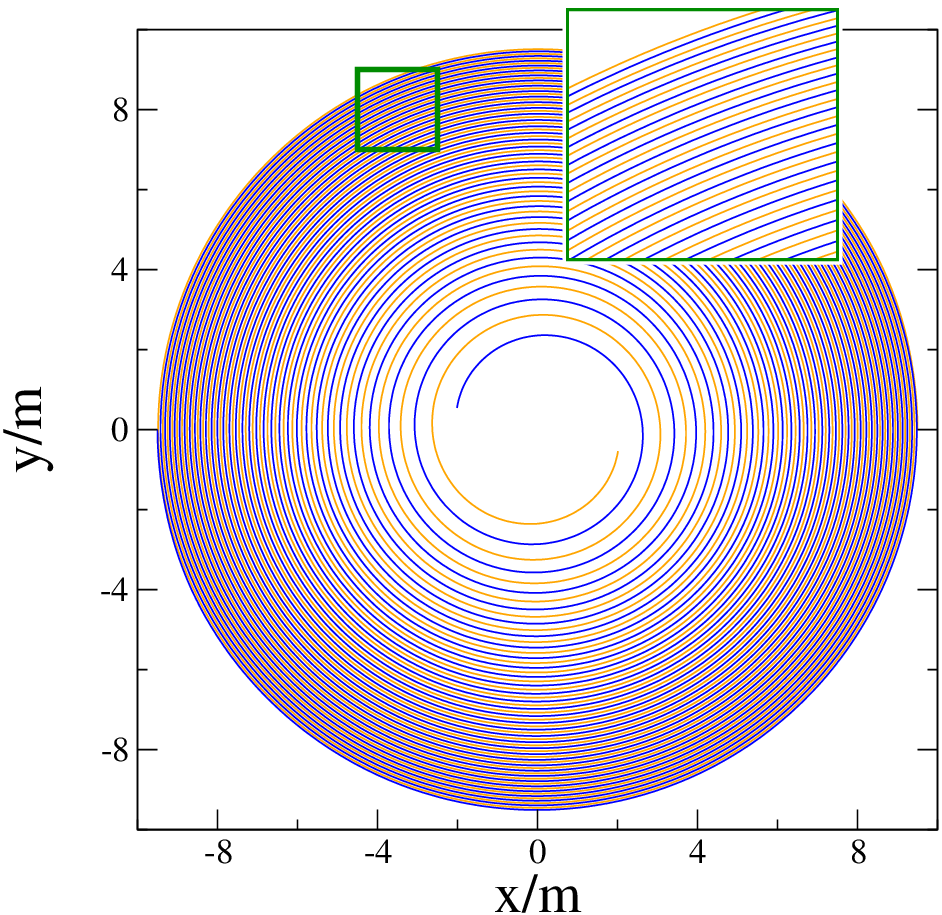}
\caption{ \label{fig:30orb_traj} The trajectories of the black holes
  for the 33-orbit numerical simulation. The blue curve
  shows the
  trajectory of one black hole and the orange curve shows the
  trajectory of the second black hole. Also shown are the individual
  apparent horizons at t=0 and at the time when the
  common apparent horizon first forms, as well as the common apparent horizon.
  }
\end{figure}

\begin{figure}
\includegraphics[width=0.9\columnwidth]{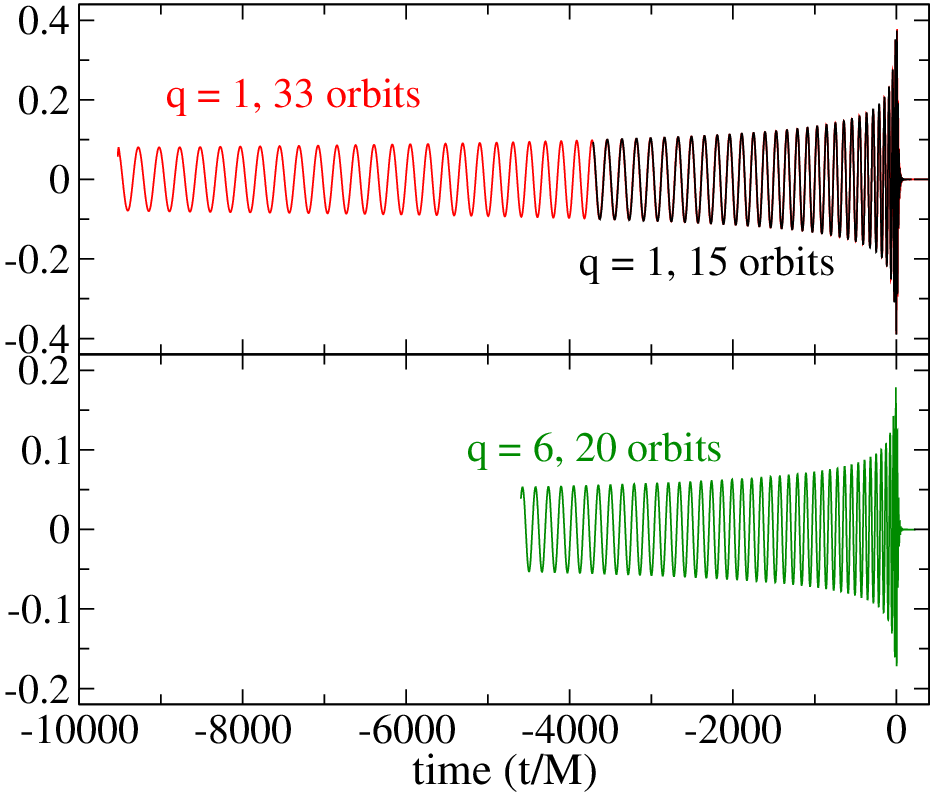}
\caption{ \label{fig:waveforms} Some of the waveforms used in this analysis.  Shown is the real part of the (2,2) mode of the equal-mass,
  non-spinning waveform with 15 orbits~\cite{Boyle2007,Scheel2009}, the new 33-orbit simulation, and the waveform for a binary with mass ratio
  $q=6$~\cite{Buchman:2012dw}. }
\end{figure}

  The numerical relativity waveforms were
  produced with the SpEC code~\cite{SpECwebsite}, a multi-domain
  pseudospectral code to solve Einstein's equations. 
  We use two simulations for the equal-mass, non-spinning binaries.  One of these two
  simulations covers 15 orbits; it was presented in
  Refs.~\cite{Boyle2007,Scheel2009} and was compared to independently
  computed equal-mass zero-spin BBH waveforms
  in~\cite{Hannam:2009hh,Garcia:2012dc}.  This 15-orbit waveform 
 was already used in the
  preceding study in MacDonald \etal~\cite{MacDonald:2011ne}.  The second
  equal-mass, non-spinning waveform covers 33 inspiral orbits.  This
  is a new simulation which is part of a larger, ongoing
  parameter-space study of binary black holes~\cite{Mroue:2012kv}. 
 This waveform was obtained
  with numerical techniques similar to those of~\cite{Buchman:2012dw}. 
The trajectories of the black holes in the 33-orbit simulation are
shown in Fig.~\ref{fig:30orb_traj}, and the corresponding
gravitational waveform in the top panel of 
Fig.~\ref{fig:waveforms}.  It is more than twice the length of the
15-orbit waveform, and will allow us to reduce the GW matching
frequency from $0.038/M$ to $\omega_m =0.025/M$ where $M = M_1 + M_2$ with $M_1$ and $M_2$ the individual masses of the black holes.  In this paper,
all frequencies $\omega$ refer to the gravitational-wave frequency that is
extracted from the $(2,2)$ mode of the gravitational waveform; at leading
order, this frequency differs by a factor of two from
the orbital frequency often used in PN calculations.

The unequal-mass waveforms of mass ratios 2, 3, 4, and 6 were
presented in detail in Buchman \etal~\cite{Buchman:2012dw}.  The
simulation with mass ratio $6$ is plotted in Fig.~\ref{fig:waveforms};
it covers about 20 orbits. The simulations with mass ratios 2, 3, and 4
are somewhat shorter and cover about 15 orbits.

\subsection{Hybridization Procedure}

The hybridization procedure used for this investigation was the same
as in our previous work (see Sec.~3.3 of~\cite{MacDonald:2011ne}): The
PN waveform, $h_\text{PN}(t)$, is time and phase shifted to match the
NR waveform, $h_\text{NR}(t)$, and they are smoothly joined
together in a GW frequency interval centered at $\omega_m$ with width
$\delta\omega$: 
\begin{equation}\label{eq:omega_match}
\omega_m-\frac{\delta\omega}{2} \le \omega \le \omega_m+\frac{\delta\omega}{2}.
\end{equation}
This translates into a matching interval $t_{\rm
  min}<t<t_{\rm max}$ because the GW frequency continuously increases during
the inspiral of the binary. As argued in~\cite{MacDonald:2011ne}, we
choose $\delta\omega = 0.1\omega_m$ because it offers a good compromise
of suppressing residual oscillations in the matching time,
while still allowing $h_\text{PN}(t)$ to be matched as closely as
possible to the beginning of $h_\text{NR}(t)$.

The PN waveform depends on a (formal) coalescence time, $t_c$,
and phase, $\Phi_c$. These two parameters are determined by minimizing
the GW phase difference in the 
matching interval $[t_\text{min}, t_\text{max}]$ as follows:
\begin{equation}
t_c', \Phi_c' = \mathrm{ \mathop{arg min}_{t_c, \Phi_c}}\int^{t_{\rm max}}_{t_{\rm min}} \big(
  \phi_{\rm PN}(t;t_c,\Phi_c) - \phi_{\rm NR} (t) \big)^2 \rm{d}t,
\end{equation}
where $t_c'$ and $\Phi_c'$ are the time and phase parameters for the
best matching between $h_\text{PN}(t)$ and $h_\text{NR}(t)$, and
$\phi(t)$ is the phase of the (2,2) mode of the gravitational
radiation.  (Because we consider only the (2,2) mode, this procedure 
is identical to time and phase shifting the PN waveform until it
has best agreement with NR as measured by the integral in Eq.~\ref{eq:omega_match})
 The hybrid waveform is then constructed in the form
\begin{equation}
h_\text{H}(t) \equiv \mathcal{F}(t) h_\text{PN}(t;t'_c,\Phi'_c) + \big[1- \mathcal{F}(t)\big]  h_\text{NR} (t), 
\end{equation}
where $\mathcal{F}(t)$ is a blending function defined as
\begin{eqnarray}
\mathcal{F}(t) \equiv 
\left\{
\begin{array}{ll}
  1, &  t < t_{\rm min} \\ 
 \cos^2\frac{\pi(t - t_{\rm min})}{2(t_{\rm max} - t_{\rm min})},\quad\quad &  t_{\rm min}
  \leq t < t_{\rm max} \\ 
  0. & t\geq t_{\rm max}  .
\end{array}
\right.\label{eq:BlendingFunction}
\end{eqnarray}

In this work, we construct all hybrids using the same procedure,
Eqs.~(\ref{eq:omega_match})--(\ref{eq:BlendingFunction}), and we vary
only the PN approximant and the matching frequency $\omega_m$.

\subsection{Quantifying errors \label{quantifying_errors}}

As in~\cite{MacDonald:2011ne}, the error measurement used to determine
the indistinguishability of two hybrid waveforms within a
gravitational wave detector's noise spectrum was $\Q$, or the
noise-weighted inner product of the difference $\delta h$ between the
two hybrids. We denote this difference $\delta h = h_\text{H1} -
h_\text{H2}$, where $h_\text{H1}$ and $h_\text{H2}$ are two hybrid
waveforms to be compared. The norm $\|\delta h\|^2 \equiv
\langle \delta h, \delta h \rangle$ is defined through the
noise-weighted inner product
\begin{equation}\label{eq:NWIP}
\NWIP{g}{h}= 2 \int_0^{\infty} df \frac{\tilde{g}^*(f)\tilde{h}(f) +
\tilde{g}(f)\tilde{h}^*(f)}{S_n(f)} \;,
\end{equation}
where $\tilde{g}(f)$ and $\tilde{h}(f)$ are the Fourier transforms of
two waveforms $g(t)$ and $h(t)$. $S_n(f)$ denotes the
(one-sided) power spectral density,
\begin{equation}
\label{eq:sn_def}
S_n(f) = 2 \int_{-\infty}^{\infty} d \tau \, e^{2 \pi i f \tau}\,
C_n(\tau)\;,\qquad f>0,
\end{equation}
where $ C_n(\tau)$ is the noise correlation matrix for zero-mean,
stationary noise. As in our previous work, we calculate
these errors using the Advanced LIGO noise curve in its high-power,
zero-detuned configuration (ZERO\_DET\_high\_P
in~\cite{Shoemaker2009}).

In order to reduce the effects of the Gibbs phenomenon in the Fourier
transforms, we apply a Planck-taper window
function~\cite{McKechan:2010kp} to the time-domain data before
computing the Fourier transform. The error is then minimized 
by a time and phase shift of one waveform relative to the other. 

Sufficient accuracy of the model waveform is guaranteed if~\cite{Lindblom2008} 
\begin{equation}\label{eq:dh-over-h-first}
\frac{\norm{\delta h}}{\norm{h}} <
\begin{cases}
  1/\rho_{\text{eff}}&\text{for parameter estimation,}\\[0.6em]
  \sqrt{2\epsilon_{\text{max}}} &\text{for event detection.}
\end{cases}
\end{equation}
Here, $\epsilon_\text{max}$ is a bound on the fractional
signal-to-noise ratio (SNR) loss while \emph{searching} for GW
signals. We follow the suggestion in~\cite{Lindblom2008} and consider
$\epsilon_\text{max} = 0.005$.  The parameter $\rho_\text{eff}$ in
Eq.~(\ref{eq:dh-over-h-first}) represents an effective SNR that
incorporates a safety factor $\varepsilon<1$~\cite{Damour:2010}, the
impact of a network of detectors, and SNR $\rho$ of the GW event. It
is defined as
\begin{equation}\label{eq:rhoeff}
\rho_{\rm eff}=\varepsilon^{-1}\; \sqrt{N}\;\rho.
\end{equation}
 where $N$ is the number of detectors. As in~\cite{MacDonald:2011ne},
 we indicate $\rho_{\rm eff} = 40$ and $\rho_{\rm eff} = 100$ in the
 plots below to cover the range of possibilities with a strong GW
 signal and a single to many detectors.
 In addition, the event-detection limit of
  Eq.~(\ref{eq:dh-over-h-first}), $\sqrt{2\epsilon_\text{max}}\approx
  0.1$, can be rewritten in terms of $\rho_{\rm eff}=10$.  We
  also indicate this bound in our figures. 

\section{RESULTS: equal-mass, non-spinning
  binaries \label{extralongNR}}

 \begin{figure}
 \includegraphics[width=0.9\columnwidth]{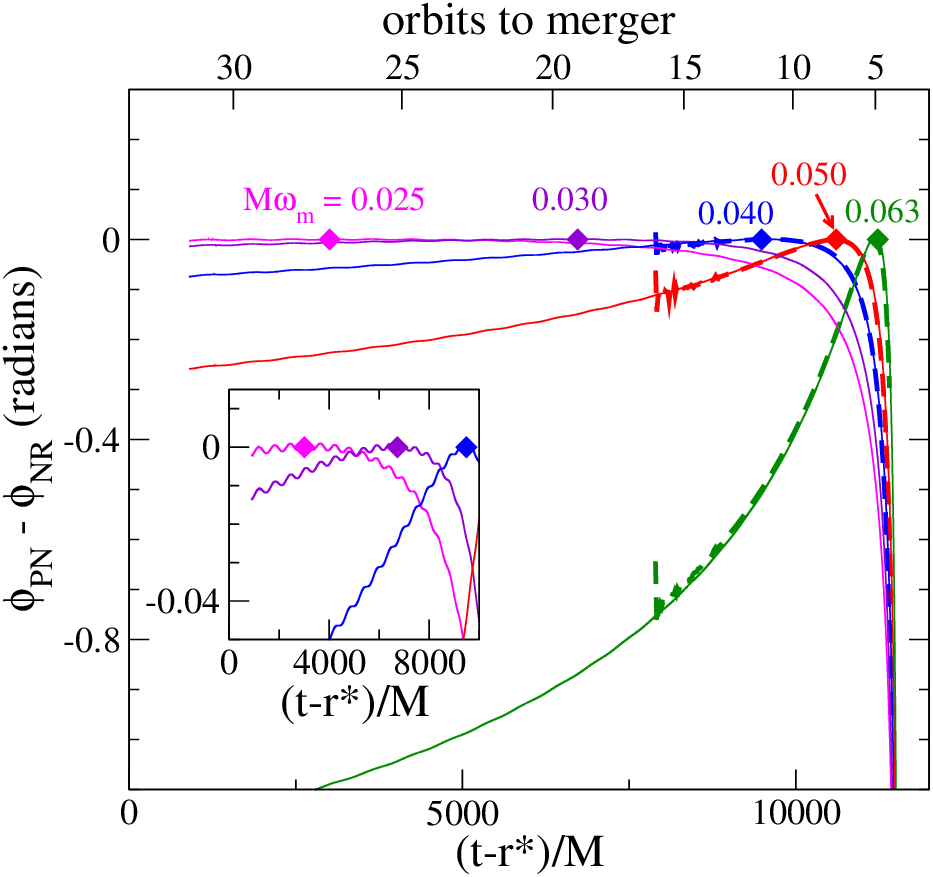}
 \caption{\label{fig:phase_err_T3} Phase error between numerical and
   post-Newtonian waveforms.  A TaylorT3 3.5/2.5PN waveform is matched
   to NR at five different GW frequencies $\omega_m$, and the
   phase differences between PN and NR are plotted.  The solid lines
   show the results obtained using the new 33-orbit numerical waveform, with
   filled diamonds indicating the location of the matching point.
   The thick dashed lines starting 16 orbits before merger
    represent the  results achieved
   in~\cite{Boyle2007}.
  }
 \end{figure}

As a first consistency check, we perform a PN comparison similar to
Boyle \etal~\cite{Boyle2007}. That work
compared a 15-orbit NR waveform to the PN approximants Taylor T1, T2,
T3, and T4. PN and NR waveforms were matched at a certain 
GW frequency $M\omega_m$, and then the differences in phases
between the two waveforms were computed as well as the relative amplitude
error. We repeat this analysis for the Taylor T3
3.5/2.5PN waveform to see if the same behaviour holds for a longer
numerical waveform. The results are shown in
Fig.~\ref{fig:phase_err_T3}. As in~\cite{Boyle2007}, we matched the 
NR and PN waveforms together at $M\omega_m = $ 0.040, 0.050, and
0.063, but also at the lower frequencies 0.025 and 0.030.

The solid lines in Fig.~\ref{fig:phase_err_T3} show the phase
error obtained when using the new 33-orbit NR waveform. The dashed
lines show the results 
from~\cite{Boyle2007}. Agreement between these two comparisons is
excellent, demonstrating that extending the SpEC simulations to larger 
numbers of orbits yields consistent results with using a shorter
waveform. This comparison validates the earlier
results~\cite{Boyle2007}, and our matching procedure, and demonstrates
consistency between the 15- and 33-orbit simulations.

\begin{figure}
\includegraphics[width=0.9\columnwidth]{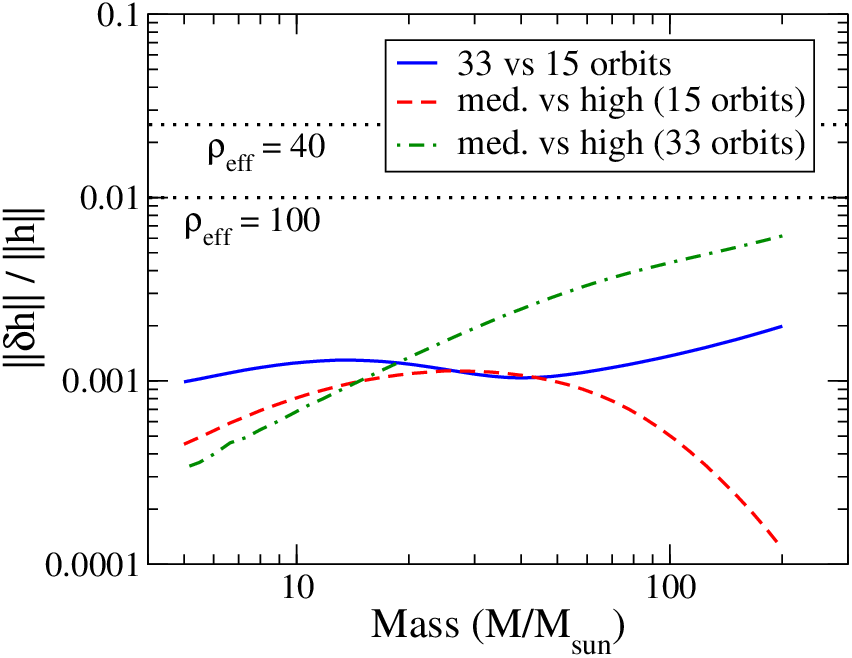}
\caption{Convergence and consistency tests between the 15- and 33- orbit
NR waveforms for q=1.  The dashed lines show truncation error
of either waveform, the solid line shows the difference between 15- and 33-orbit waveforms.  For all three curves, hybrids were constructed with the same Taylor approximant at the same matching frequency.  $\Q$ was then evaluated 
as a function of mass $M$ using the advanced LIGO noise curve. \label{fig:30orb_error}}
\end{figure}

Continuing the consistency tests between the 
15-orbit waveform~\cite{Boyle2007,Scheel2009} and the new 33-orbit
waveform presented in Sec.~\ref{NumericalWaveforms}, 
Fig.~\ref{fig:30orb_error} shows the error $\Q$
between hybrids created with the two numerical waveforms matched at the same hybridization frequency
$M\omega_m = 0.042$ as a function of total mass, as well as the error
between the high and medium resolutions for each waveform. This figure
shows first that the error between hybrids created with 
the longer and shorter NR waveforms is comparable to the numerical
error of either simulation. Numerical errors are much smaller than the
error bound for $\rho_\text{eff} = 100$, and so we shall therefore
disregard numerical errors.

\begin{figure}
\includegraphics[width=0.9\columnwidth]{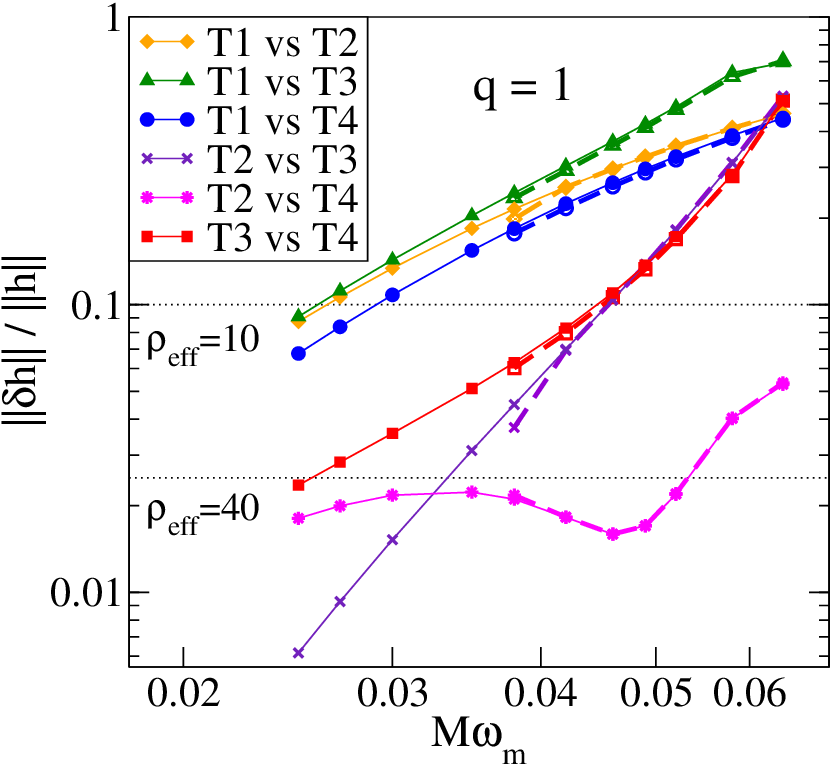}
\caption{\label{fig:T1vsT2vsT3vsT4_loglog_q1} 
Error of hybrid waveforms as a function of matching frequency for
mass ratio $q=1$.
The solid lines show
  $\Q$ for hybrids constructed with different PN approximants and with
  the new 33-orbit NR waveform. The dashed lines show the previous
  results which use a 15-orbit NR waveform. These errors are all for
  binary black-hole systems with a total mass of $20 M_\odot$.}
\end{figure}

Having established the accuracy of the 33-orbit waveform, let
  us now assess the quality of PN-NR hybrid waveforms at the lower
  matching frequencies that are made accessible by this NR waveform.
  We choose several matching frequencies in the interval $0.025\le
  M\omega_m\le 0.063$, and for each of these, we construct hybrid
  waveforms with Taylor T1, T2, T3, and T4.  We then compute pairwise
  differences $\Q$ at a fixed total mass $M=20M_\odot$
  using the ZERO\_DET\_HIGH\_P LIGO noise-curve~\cite{Shoemaker2009}.
  The results of this comparison are shown as the solid lines in
  Fig.~\ref{fig:T1vsT2vsT3vsT4_loglog_q1}.  Differences decrease with
  decreasing matching frequency (i.e., when switching from PN to NR
  earlier), as one would expect.  The thick dashed lines represent the
  same comparison performed with the 15-orbit waveform, i.e.,  the
  precise data already presented in Ref.~\cite{MacDonald:2011ne}.  The
  comparisons agree for $M\omega_m\ge 0.038$, where both NR waveforms
  are available.  Furthermore, the long waveform continues the trend
  set by the earlier, shorter waveform without surprises.  This
  indicates that extrapolating errors in our analyses to lower
  matching frequencies is indeed valid.  It also indicates that there
  is no unexpected behaviour of PN or NR in the newly accessible
  GW frequency interval $0.025\le M\omega\le 0.038$, with PN and NR
  converging toward each other.

\section{Unequal-mass black holes \label{unequalmass}}

\begin{figure}
\includegraphics[width=0.9\columnwidth]{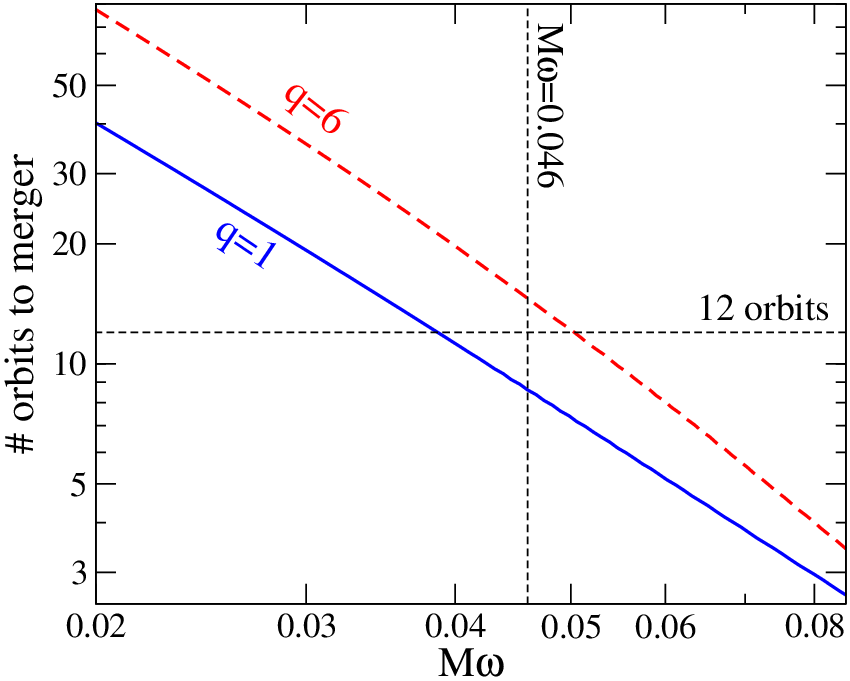}
\caption{\label{fig:OrbitsVsOmega} The relationship between the number
of orbits before merger and the gravitational wave frequency for
mass ratios $q=1$ and $q=6$.} 
\end{figure}

Now let us consider binary black-hole systems with mass ratio $q =
M_1/M_2 = 2,3,4,6$.  As $q$ increases,
  the inspiral proceeds more slowly, in
  proportion to the symmetric mass ratio $\nu=M_1M_2/M^2$ (see
  e.g., post-Newtonian expansions~\cite{Blanchet2006}).  This is
  illustrated by Fig.~\ref{fig:OrbitsVsOmega}, which shows the number
  of inspiral orbits to merger as a function of gravitational wave
  frequency. This figure was created by plotting the number of orbits
  prior to the maximum amplitude of hybrid waveforms matched at the
  earliest possible GW frequency ($\omega_m = 0.025$ for $q=1$ and 0.042
  for $q=6$) with a Taylor T4 waveform against $\omega$.

  Starting from the same GW frequency (e.g., $M\omega=0.046$), 
  the $q=6$ binary proceeds through roughly twice as many orbits.
  Conversely, the same number of orbits to merger (e.g., $N=12$) occurs
  at higher frequency for $q=6$ than for $q=1$.  The dashed lines
  indicate a matching frequency of $M\omega_m = 0.046$ and 12 orbits;
  we will use these two reference values in subsequent comparisons.
  Because unequal-mass binaries spend more orbits in the strong field
  regime, we would expect that with increasing mass ratio $q$, hybrids
  need to be matched a {\em larger} number of orbits before merger
  than for $q=1$.  The next sections will quantify this expectation.

Before proceeding, it is important to note that the error
  between hybrids constructed with different PN approximants is much
  higher than the numerical error of the NR waveforms. We performed an
  analysis similar to that in Fig.~\ref{fig:30orb_error}, where the
  highest resolution NR waveform was compared to the medium resolution
  waveform for all mass ratios used in this paper. In all cases, the
  numerical error is about an order of magnitude smaller than the
  PN error in hybrid waveforms.

\subsection{Phase errors between PN and NR \label{phase_errors}}

Let us start with investigating the phase difference between NR and PN
with measures that are independent of the LIGO noise spectrum. We
calculate the \emph{accumulated phase difference} between PN and NR
waveforms. Following Hannam \etal~\cite{HannamEtAl:2010}, we
match PN and NR at $M\omega_m = 0.1$ and then calculate
their phase difference at a certain time before
this matching point. 

\begin{figure}
   \includegraphics[width=0.9\columnwidth]{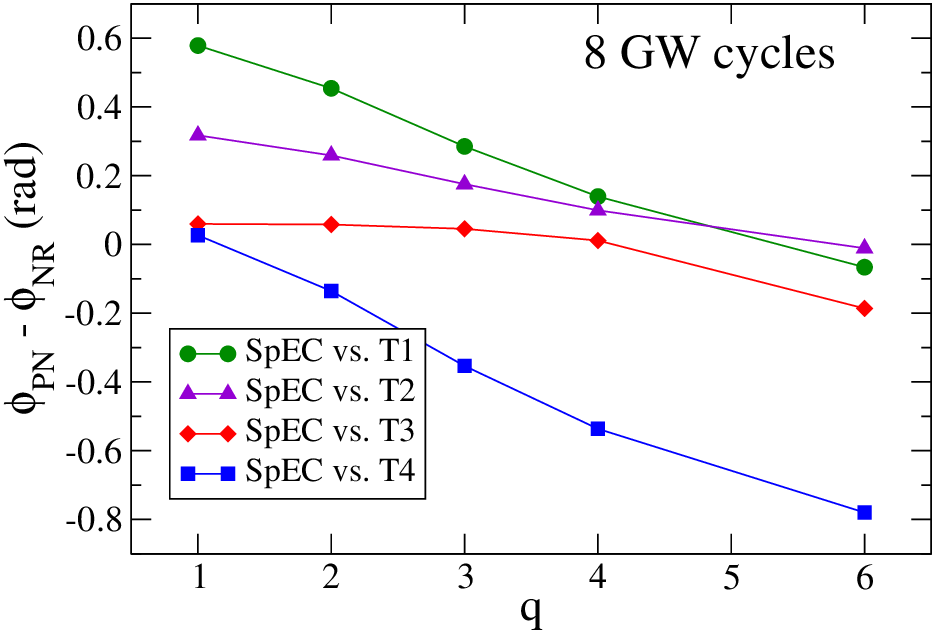}\\[2em]
   \includegraphics[width=0.9\columnwidth]{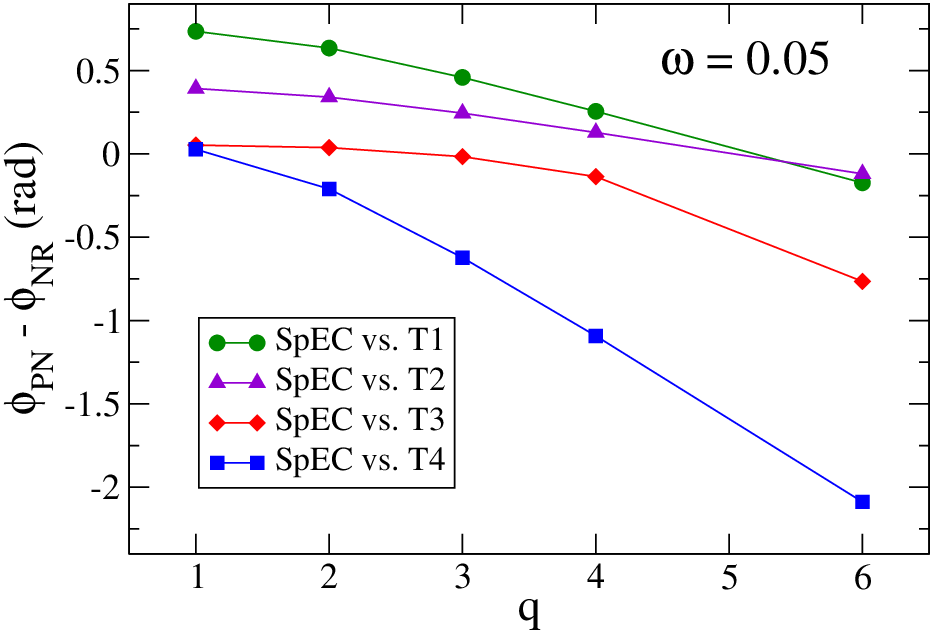}
   \caption{\label{fig:AccumDiff} Accumulated phase difference between
     numerical and post-Newtonian waveforms for Taylor T1, T2, T3, and
     T4.  PN and NR are matched at GW frequency $M\omega_m=0.1$,
     phase differences are then computed eight GW cycles earlier (top
     panel) or at GW frequency $M\omega=0.05$ (bottom panel).  The
     Taylor T3 waveform used in this comparison is 3.0 PN order in
     phase.  }
 \end{figure}

We perform this computation for TaylorT1, TaylorT2,
TaylorT3, and TaylorT4. (The TaylorT3 waveform was only
calculated to 3.0PN order in phase since the 3.5PN order waveform does
not reach $M\omega = 0.1$.) The results are shown in
Fig.~\ref{fig:AccumDiff}. The top panel shows the phase difference
between PN and NR at 8 GW cycles before $M\omega_m = 0.1$ (equivalent
to Fig. 8 of~\cite{HannamEtAl:2010}), and the 
bottom panel shows the phase difference at $M\omega= 0.05$ for PN and
NR waveforms matched in phase at $M\omega_m = 0.1$. Both panels show
similar trends. In the lower panel, the phase difference increases
more rapidly with increasing mass ratio because the number of orbits
within the comparison increases with increasing mass ratio (7.4 orbits
for $q=1$ versus 12.2 orbits for $q=6$). 

Our findings disagree with the results of Hannam 
\etal~\cite{HannamEtAl:2010}, despite following the identical
  comparison protocol.  Figure~8 of~\cite{HannamEtAl:2010} shows a
  roughly constant
  phase difference between TaylorT1 and the numerical simulation of $\sim
  0.5$ radians, for all considered mass ratios ($q=1,2,3,4$), whereas we
  find a steadily declining phase difference reaching zero near
  $q\sim5$.  Similar trends hold for TaylorT4: 
  Ref.~\cite{HannamEtAl:2010} reports phase differences $\sim -0.1$ radians
  for mass ratios $q=1,2,3,4$.  We see a very small phase difference
  at mass ratio $q=1$, and steadily increasing phase differences
  at larger mass ratios. These discrepancies are caused by an error in
  the numerical setup of the simulations used in
  Ref.~\cite{HannamEtAl:2010}\footnote{M. Hannam, S. Husa, and
    M. P\"urrer (private communication, October 29, 2012)}.

 Figure~\ref{fig:AccumDiff} shows that one cannot assume that any
 single PN approximant will be suitable for all of parameter
 space. Phase errors between PN and NR change dramatically with
 increasing mass ratio. The agreement between TaylorT4 and NR
 waveforms in the case of equal-mass non-spinning binaries, for
 example, is purely coincidental. As the mass ratio of the binary
 increases, this phase error becomes much larger for TaylorT4, and in
 fact becomes smaller for TaylorT1 and TaylorT2.  Because the
   Taylor approximants differ only in higher order post-Newtonian
   terms, the spread between them can be taken as a measure of the
   post-Newtonian truncation error.  For the comparison in the top
   panel, this indicates a post-Newtonian truncation error of $\sim
   0.5$ radians for $q=1$ increasing to $\sim 1$ radian for $q=6$.  Within
   this (admittedly large) truncation error, all four
   Taylor approximants are consistent with the numerical data
   (i.e., consistent with zero phase difference).

\subsection{Hybrid errors}

\begin{figure}
\includegraphics[width=0.9\columnwidth]{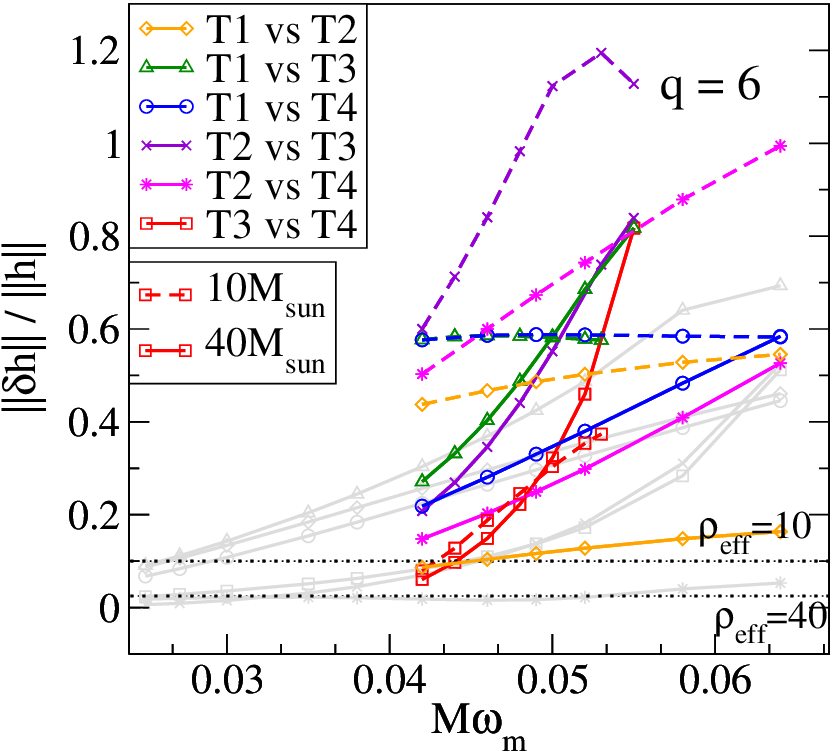}
\caption{Error of hybrid waveforms as a function of matching frequency for
mass ratio $q=6$. Plotted is the value of
  $\Q$ at 10M$_\odot$ and 40M$_\odot$ as a function of $\omega_m$
  for a binary black-hole system with a mass ratio $q = 6$. The data from
  Fig.~\ref{fig:T1vsT2vsT3vsT4_loglog_q1} are 
  plotted in grey in the background for
  reference. \label{fig:T1vsT2vsT3vsT4_q6}}
\end{figure}

We now repeat the analysis of
Fig.~\ref{fig:T1vsT2vsT3vsT4_loglog_q1} for binaries
with higher mass ratios: We hybridize TaylorT[1,2,3,4] at several
matching frequencies $\omega_m$. At each $\omega_m$, we compute
differences $\Q$ between all six pairs of PN approximants and plot
these differences as a function of
$\omega_m$. The results are shown in Fig.~\ref{fig:T1vsT2vsT3vsT4_q6} 
for mass ratio $q = 6$ and for total masses 10$M_\odot$
and 40$M_\odot$. $M=10M_\odot$ represents a binary with
component masses 1.67$M_\odot$ and 8.33$M_\odot$, which we shall take
as an approximation of a black-hole--neutron-star (BH-NS) binary. $M=40M_\odot$
results in component masses 6.67$M_\odot$ and 33.337$M_\odot$, a BBH
system where the smaller black hole is consistent with 
known black-hole masses. We chose these masses because they are more
astrophysically probable for this mass ratio than the total mass of
20$M_\odot$ that we used for $q =1$. The coloured lines represent
the differences between hybrids as just described. To ease the
comparison with the $q = 1$ results of
Fig.~\ref{fig:T1vsT2vsT3vsT4_loglog_q1}, we duplicate those data into
Fig.~\ref{fig:T1vsT2vsT3vsT4_q6} as the grey lines in the
background.

One notices immediately two differences between $q = 1$ and $q = 6$:
(1) At the same matching frequency, $q = 6$ results in larger
differences. This might be caused by the larger number of orbits that
the $q = 6$ binary spends at high frequency
cf. Fig.~\ref{fig:OrbitsVsOmega}. (2) The $q = 6$ comparison covers
only comparatively high matching frequencies $M\omega_m \geq 0.042$,
whereas the $q = 1$ comparison reaches much lower frequencies. This
originates in the slower inspiral of higher mass ratios (i.e., longer
time to merger from the same starting frequency) and the higher
computational cost of high mass ratio simulations.

The difference between the TaylorT1 and TaylorT2 hybrids is
particularly small in Fig.~\ref{fig:T1vsT2vsT3vsT4_q6}. This is
consistent with Fig.~\ref{fig:AccumDiff}, where for $q =
6$, the differences between PN and NR are similar for these two
approximants. 

\begin{figure}
\includegraphics[width=0.96\columnwidth]{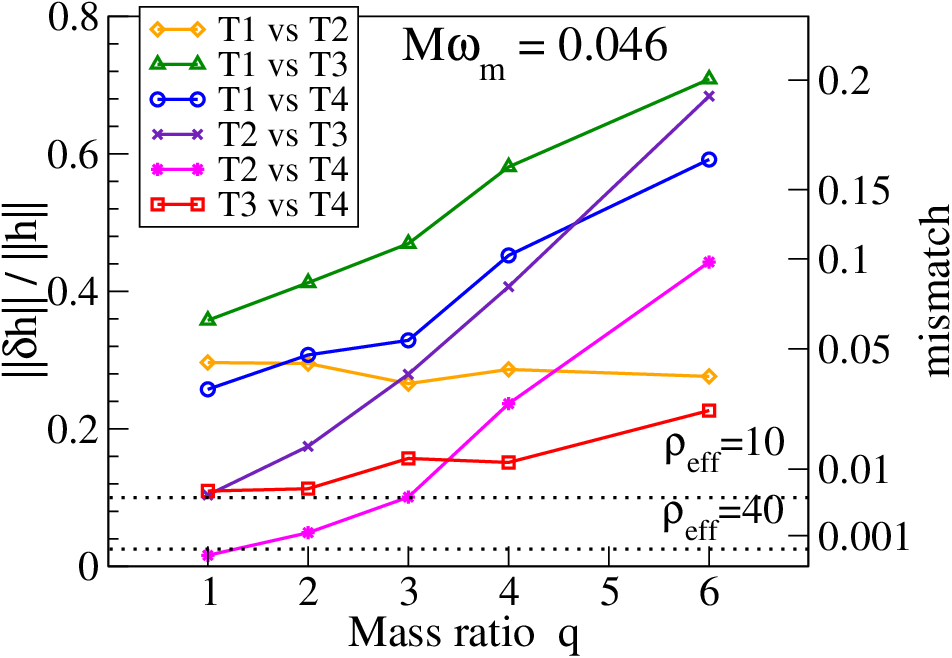}
\includegraphics[width=0.96\columnwidth]{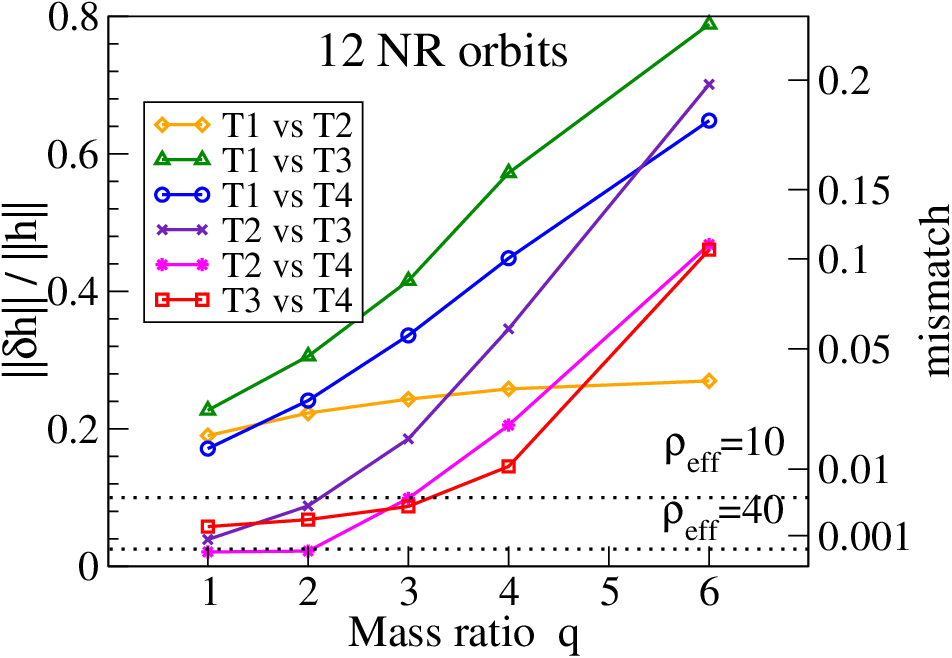}
\caption{Error of hybrid waveforms as a function of mass ratio $q$. 
  PN and NR is matched at frequency $M\omega_m=0.046$ (top panel) or
  12 orbits before merger (bottom panel).  Pairwise differences between
hybrids in either category are computed at total mass $M=20M_\odot$.
  \label{fig:T1vsT2vsT3vsT4_omega046}}
\end{figure}

Let us now investigate the dependence on $q$ in more detail. We
compute the differences $\Q$ for all mass ratios at the same matching
frequency $M\omega_m=0.046$ (the data for $q = 1$ and 6 can be read
off of Figs.~\ref{fig:T1vsT2vsT3vsT4_loglog_q1}
and~\ref{fig:T1vsT2vsT3vsT4_q6}, respectively). The upper panel of
Fig.~\ref{fig:T1vsT2vsT3vsT4_omega046} shows these differences as a
function of $q$. Once again, it is evident that differences increase
with mass ratio.

Length requirements for NR waveforms are often phrased in the
convenient unit ``number of NR orbits''. To place this notion in
context, we match PN+NR 12 orbits before merger, compute differences
$\Q$, and plot these in the lower panel of
Fig.~\ref{fig:T1vsT2vsT3vsT4_omega046}. When matching a fixed number
of orbits before merger, $\Q$ increases even more steeply with $q$,
because the matching frequency increases with $q$,
cf. Fig.~\ref{fig:OrbitsVsOmega}.

Both Figs.~\ref{fig:AccumDiff}
and~\ref{fig:T1vsT2vsT3vsT4_omega046} show that the error between
hybrids using different PN approximants increases with mass ratio.
Therefore, with increasing mass ratio
binaries, NR waveforms will have to be longer to create hybrids
of similar quality.  For mass ratio 6, one might estimate from
  Fig.~\ref{fig:T1vsT2vsT3vsT4_q6} that total mass $M=40M_\odot$
  requires a  matching frequency of $M\omega_m\approx 0.03$.  The
  matching interval Eq.~(\ref{eq:omega_match}) would then extend to a
  lower frequency $M\omega=0.0285$ requiring about 40 orbits covered by NR. 
The needed matching frequency (and thus the number of orbits) depends on the
total mass considered; for $M=10M_\odot$, convergence of the errors with decreasing $M\omega_m$ is not yet apparent (see the $10M_\odot$ curves in
Figure~\ref{fig:T1vsT2vsT3vsT4_q6}), 
so the NR waveform is too short to even estimate how long it should be.  
This indicates that 
 BH-NS systems may very well place the most stringent
  requirements on NR simulations.  (A proper treatment of BH-NS
  systems, of course, will also require to simulate the neutron star
  directly, including its tides and other effects arising from
  micro-physics.  Such a simulation would be yet more challenging than
  our approach of using the 
  easier BBH system as a proxy.)

It is no surprise that the error for hybrids matched at a certain
$M\omega_m$ or at a fixed number of orbits before merger would
increase with mass ratio since the number of orbits spent in the strong field regime increases with mass ratio. 
 
 \section{Higher-order Post-Newtonian \label{4PNerror}}
 
The primary source of error lies in the truncation error of the PN
 approximants. Work is currently being done to calculate PN to 4th
 order~\cite{Jaranowski:2012eb,Foffa:2012rn}, therefore, an interesting question arises:
 how much might higher PN orders improve the accuracy of PN+NR
 hybrid waveforms? To address this question we consider the TaylorT4
 approximant, where the phase evolution is 
 determined by a single Taylor series~\cite{Buonanno:2002fy}:  
 \begin{eqnarray} \label{eq:T4}
  \frac{dx}{dt} &=& \frac{64c^3\nu}{5GM}x^5 \Bigg\{ 1 -
  \left(\frac{743}{336} + \frac{11}{4}\nu \right)x  + 4\pi x^{3/2}
   \nonumber \\
&+&   \left(\frac{34103}{18144} + \frac{13661}{2016}\nu +
  \frac{59}{18}\nu^2 \right)x^2  -  \left( \frac{4159}{672}
\right. \nonumber \\
&+& \left. \frac{189}{8}\nu \right)\pi x^{5/2} + \left[
  \frac{16447322263}{139708800} + \frac{16}{3}\pi^2 -
  \frac{1712}{105}\gamma \right. \nonumber \\
&+& \left( \frac{451}{48}\pi^2 - \frac{56198689}{217728} \right)\nu +
  \frac{541}{896}\nu^2 - \frac{5605}{2592}\nu^3 \nonumber \\
 &-&  \left. \frac{856}{105}\ln{(16x)} \right]x^3 - \left(
   \frac{4415}{4032} - \frac{358675}{6048}\nu  \right. \nonumber \\
&-&  \left. \frac{91495}{1512}\nu^2 \right)\pi x^{7/2}
   + A_\text{4}x^4 + A_\text{4.5}x^{9/2} + A_\text{5}x^5 \Bigg\}.
 \end{eqnarray}
Here, $\nu=M_1M_2/M^2$ denotes the symmetric mass ratio, $\gamma$ is
Euler's constant, $c$ is the speed of light, $G$ is the gravitational
constant, and $x=v^2/c^2$. 

\begin{figure}
\includegraphics[width=0.9\columnwidth]{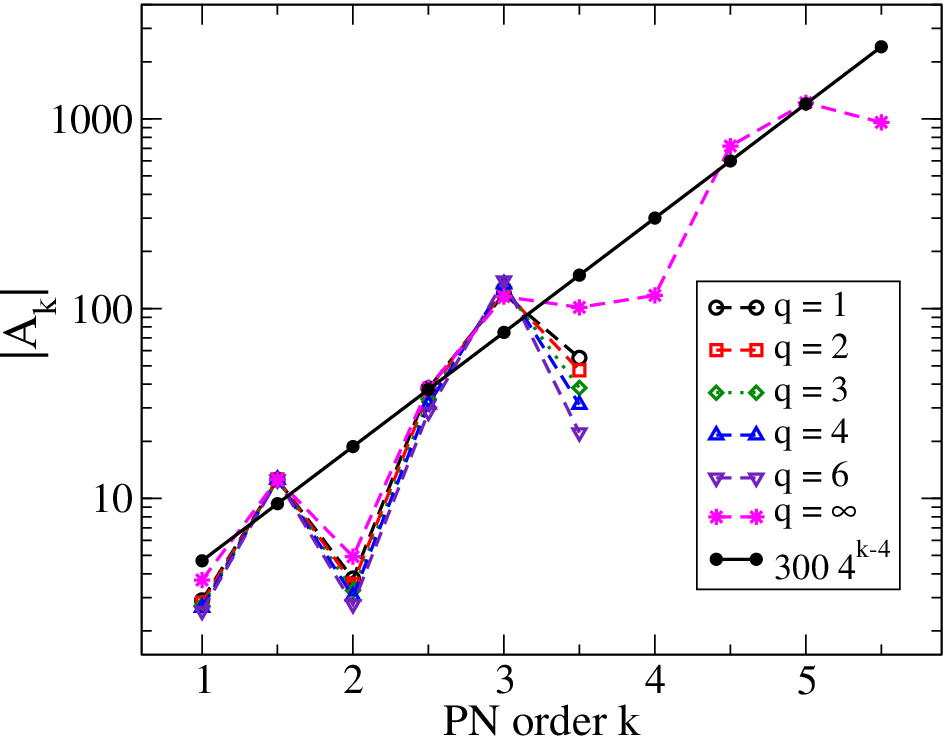}
\caption{The magnitude of the PN coefficients for Taylor T4 as a
  function of PN order.  Shown are different mass ratios and the
  test-mass limit.  The straight black line shows our assumed coefficients
  for estimating the accuracy of 4PN and 4.5PN hybrids.
 \label{fig:A_PN_vs_PNorder}}
\end{figure}

In Eq.~\ref{eq:T4}, we have included  the
terms $A_\text{4}$, $A_{4.5}$, and $A_\text{5}$ at
4th through 5th PN order.  These terms are currently unknown, 
but trends in the known coefficients can 
 be used to approximate
the magnitude of the unknown coefficients.
Figure~\ref{fig:A_PN_vs_PNorder} plots the known coefficients $A_k$ vs. PN order $k$ for mass ratios $q=1,2,3,4,6$.  The figure also shows the 
coefficients for the test-mass limit $\nu\to 0$, which are 
known up to 5.5-th PN order~\cite{Boyle:2008}.   The coefficients
$A_k$ behave rather erratically, but overall, they seem to be
exponentially increasing, as indicated by the black line with filled
circles. This black line represents $A_4 = 300$, with a
doubling of the coefficients with each increase in PN order (and
halving with each decrease). Therefore, we shall take the
unknown coefficients to be $A_4=300$, $A_{4.5}=600$ and $A_5=1200$.
These values are of course not the correct ones, but are indicative of 
the expected magnitudes of these coefficients.  Hence, we will not be able to
compute the correct 4PN (and higher) PN waveforms, but merely estimate the
errors in those waveforms, were the coefficients known.

\begin{figure}
\includegraphics[width=0.9\columnwidth]{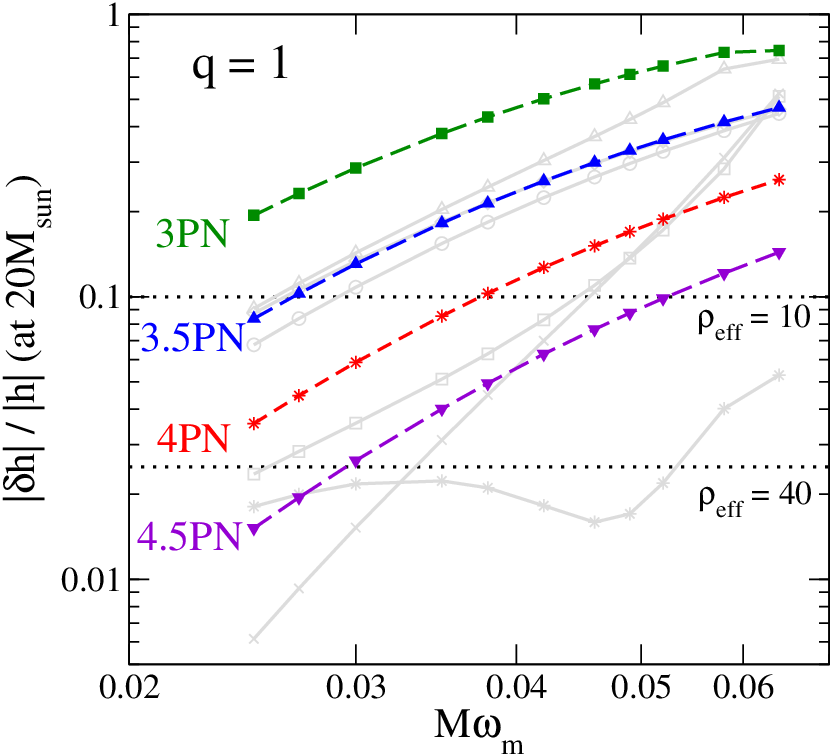}
\caption{\label{fig:dtbiasVSomegaT4} 
Estimated error of hybrid waveforms if PN up to the specified order were known. 
Shown is the equal-mass case.  The line ``3.5PN'' represents the 
error in the presently available PN waveforms versus
an estimated 4PN term; see the text for details.  
The grey lines show the more exhaustive 
analysis from Fig.~\ref{fig:T1vsT2vsT3vsT4_loglog_q1}, which is consistent
with the ``3.5PN'' line obtained with our alternative estimation procedure.} 
\end{figure}

Assuming $A_\text{4}=300$, we can estimate the truncation error of
3.5PN TaylorT4 by computing $\Q$ between the standard TaylorT4 (with
$A_k=0$ for $k \geq 4$) and a modified TaylorT4 with
$A_\text{4}=300$. This comparison is shown as the blue curve in
Fig.~\ref{fig:dtbiasVSomegaT4} which is labeled ``3.5PN'' (the
  label indicates that this is an estimate of the error of 3.5PN
  order).  This new estimate of the 3.5PN truncation error should of
  course be consistent with our earlier estimates shown in
  Fig.~\ref{fig:T1vsT2vsT3vsT4_loglog_q1}.  To demonstrate this
  consistency, we include the data of
  Fig.~\ref{fig:T1vsT2vsT3vsT4_loglog_q1} as the greyed out lines in
  the background of Fig.~\ref{fig:dtbiasVSomegaT4}.  Indeed, the new
  estimate (``3.5PN'') follows closely the trends of the
  more exhaustive study, lending confidence in this approach.

Repeating this procedure at next higher PN order will now result in an
error estimate of 4PN (were it known).  Thus we compare TaylorT4
hybrids with and without the $A_\text{4.5} = 600$ term. This results in
the red line labeled ``4PN'' in Fig.~\ref{fig:dtbiasVSomegaT4}. We
can go one PN order further, and include an $A_\text{5} $ term
(resulting in the line labeled ``4.5PN'').  We can also remove
the (known) 3.5 PN term, to estimate the 3PN truncation error if 3.5PN
were not known (the curve labeled ``3PN''; this curve compares $A_\text{3.5}=0$
with $A_\text{3.5}=150$).

\begin{figure}
\includegraphics[width=0.9\columnwidth]{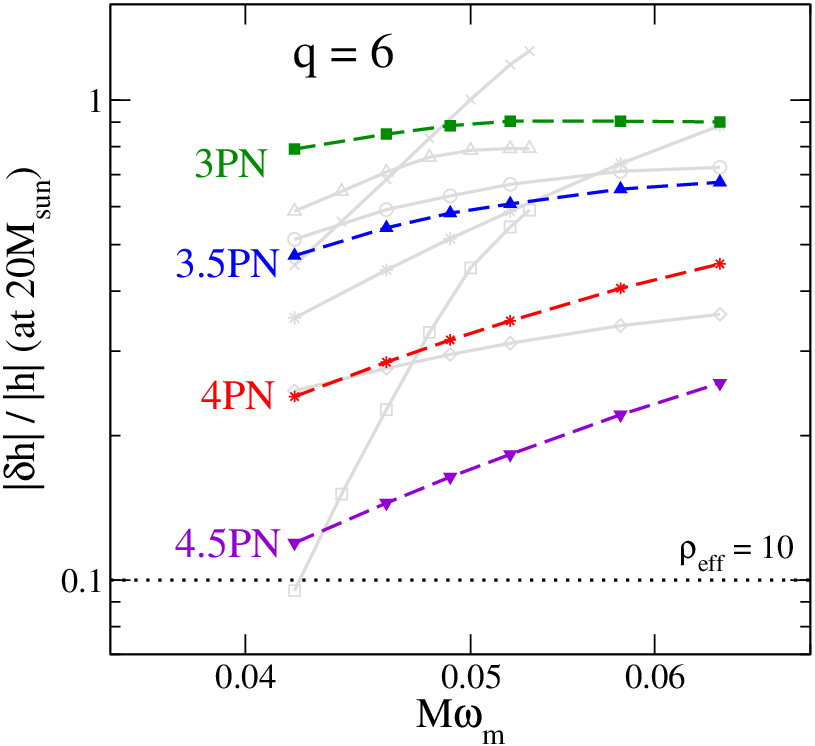}
\caption{\label{fig:dtbiasVSomegaT4_q6}
Estimated error of hybrid waveforms if PN up to the specified order were known. 
Shown is the mass ratio $q=6$ case.  The line ``3.5PN'' represents the
error in the presently available PN waveforms. The grey lines show the more
exhaustive analysis from Fig.~\ref{fig:T1vsT2vsT3vsT4_q6}, but at a
total mass of $20 M_\odot$. Compare to
Fig.~\ref{fig:dtbiasVSomegaT4}. }  
\end{figure}

\begin{figure}
\includegraphics[width=0.9\columnwidth]{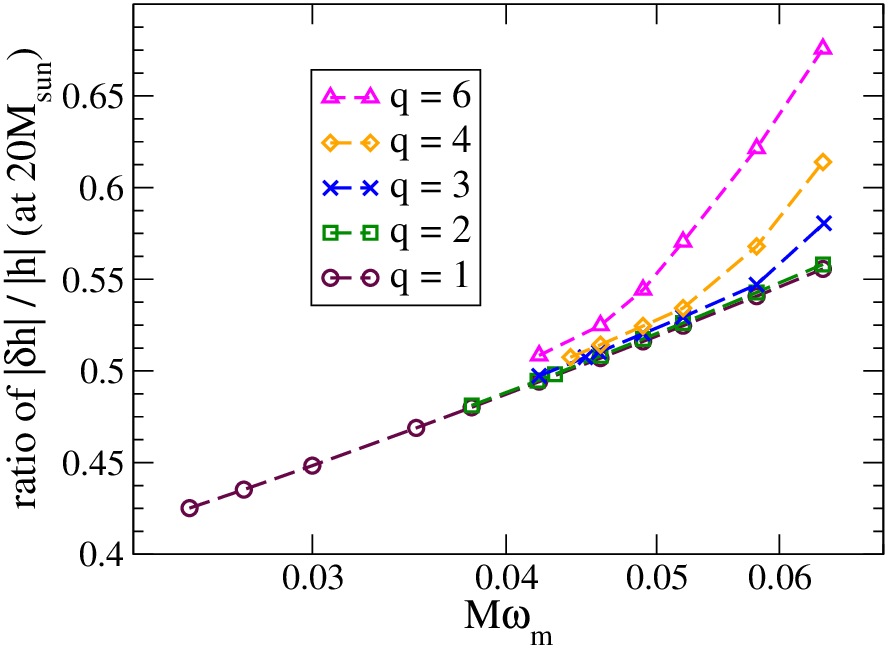}
\caption{ \label{fig:RatioOfErrorsVsOmega_diffq} 
Estimated reduction in error of hybrid waveforms constructed using 4PN
relative to those constructed with 3.5PN. 
This reduction is shown as a function of matching frequency for
different mass ratios.  The $q=1$ line is
the ratio between the ``4PN'' and ``3.5PN'' curves in
Fig.~\ref{fig:dtbiasVSomegaT4}. }
\end{figure}

We repeated this analysis for the mass ratios $q=2,3,4,$ and 6.
  The results are similar to the $q\!=\!1$ calculation of
  Fig.~\ref{fig:dtbiasVSomegaT4}; as an example,
  Fig.~\ref{fig:dtbiasVSomegaT4_q6} presents the analogous calculation
  for $q\! =\! 6$.  In Fig.~\ref{fig:dtbiasVSomegaT4}
and~\ref{fig:dtbiasVSomegaT4_q6}, a clear pattern emerges: Each additional PN order
reduces $\Q$ by approximately a factor of $\sim 2$.  When matched
at low frequency $M\omega_m$ (where PN is more accurate) the reduction
in error is somewhat faster than when matching at high frequency.
To make these statements quantitative,
Figure~\ref{fig:RatioOfErrorsVsOmega_diffq} plots the ratio of the
``3.5PN'' and the ``4PN'' curves in Fig.~\ref{fig:dtbiasVSomegaT4}
and~\ref{fig:dtbiasVSomegaT4_q6}.  It also shows data for the 
remaining mass ratios ($q = 2,3,4$).   This ratio is $0.5$ at
$M\omega_m\approx 0.04$ and drops to $0.42$ at the lowest accessible
matching frequency $M\omega_m=0.025$.  The gain of higher-order PN is
approximately independent of mass ratio, except for high matching
frequencies and high mass ratios; in this regime errors are so large
that the asymptotic trends for small errors/early matching are
masked. 

\subsection{3PN Hybrids}

\begin{figure}
\includegraphics[width=0.9\columnwidth]{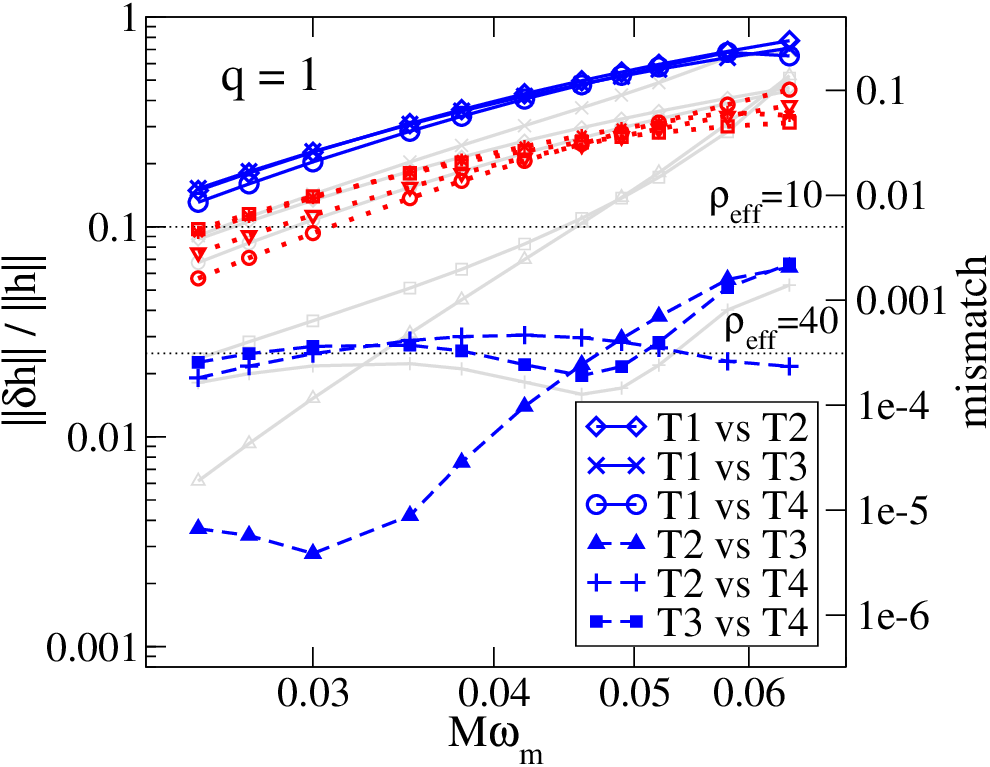}
\caption{\label{fig:T1vsT2vsT3vsT4_30_30_q1}
Error analysis of PN-NR hybrids using only 3PN information.
The solid and dashed blue lines repeat the analysis of Fig.~\ref{fig:T1vsT2vsT3vsT4_loglog_q1} at 3PN order.  
 The
  dotted red lines show $\Q$ for hybrids using 3PN waveforms
  compared to one using Taylor T4 at 3.5PN (the circle is T1, the star
  T2, the square T3 and the downward triangle is T4).   For reference, the
 data of Fig.~\ref{fig:T1vsT2vsT3vsT4_loglog_q1} are
 included in grey in the background.
}
\end{figure}

To further quantify the reliability of our
estimate of 4PN improvements, let us apply our analysis to 3PN
waveforms. The 3PN truncation error estimate based on modifying the
3.5PN $A_\text{3.5}$ coefficient is already plotted in
Figs.~\ref{fig:dtbiasVSomegaT4} and~\ref{fig:dtbiasVSomegaT4_q6}. What
remains is the equivalent of Fig.~\ref{fig:T1vsT2vsT3vsT4_loglog_q1}
at 3PN order: We prepare TaylorT[1,2,3,4] hybrids at 3PN order,
and compute their pairwise differences. This results in the thick
solid blue lines of Fig.~\ref{fig:T1vsT2vsT3vsT4_30_30_q1}. Compared
to the 3.5PN comparison of Fig.~\ref{fig:T1vsT2vsT3vsT4_loglog_q1}
(replicated in the grey lines in
Fig.~\ref{fig:T1vsT2vsT3vsT4_30_30_q1}), the error of 3PN is indeed
somewhat larger than for 3.5PN. The change in error 3PN vs. 3.5PN is
smaller than the factor of 2 we would have predicted from
Fig.~\ref{fig:dtbiasVSomegaT4}, presumably because the actual
$A_\text{3.5}$ coefficient is smaller than the assumed
$A_\text{3.5} = 150$ of Fig.~\ref{fig:A_PN_vs_PNorder}.

Fig.~\ref{fig:T1vsT2vsT3vsT4_30_30_q1} offers another cautionary
lesson:  Consider the differences between hybrids using the
3PN Taylor T2,T3,T4 approximants.  There are three such differences, 
plotted with blue dashed lines in Fig.~\ref{fig:T1vsT2vsT3vsT4_30_30_q1}.
The differences between these three hybrids
are surprisingly small, with $\Q \leq 0.07$, even when
hybridized at the very large matching frequency $M\omega_m =
0.065$. If one had used only these three approximants, and if only
3PN were available, one would have concluded that the resulting
hybrids are good for Advanced LIGO purposes, and that short NR
simulations are sufficient.

But this conclusion would have been entirely incorrect! The dotted red
lines in Fig.~\ref{fig:T1vsT2vsT3vsT4_30_30_q1} show the difference
between 3PN hybrids with the 3.5PN TaylorT4 hybrid. These differences
are an order of magnitude larger than the 3PN internal TaylorT2, T3, and T4
differences. At 3PN, TaylorT2, T3, and T4 are very similar, and they
are all three off by the same large amount.  The 3PN TaylorT1
  comparisons reveal this effect: As shown by the solid blue lines in
  Fig.~\ref{fig:T1vsT2vsT3vsT4_30_30_q1}, differences between TaylorT1
  hybrids and the other three hybrids are all very large.  As this
  analysis demonstrates, these large differences are not caused by a
  ``deficiency'' of TaylorT1, but rather by coincidentally similar
  deviations from the true waveform in the TaylorT2,3,4 approximants.
The same effect at 3.5PN order is apparent in Fig. 5: At low matching
frequencies, the pairwise differences between TaylorT2,3,4 are
significantly smaller than the differences relative to TaylorT1.
Therefore, it is important to investigate many different
approximants, and not to discount lone outliers as in
Fig.~\ref{fig:T1vsT2vsT3vsT4_30_30_q1}.

 \section{Discussion}

In this paper, we have presented an analysis of errors that affect
hybrid gravitational waveforms for a range of mass ratios to assess
their suitability for parameter estimation with Advanced
LIGO. We have also estimated by how much these errors would be reduced
if PN were known to a higher order.

In the case of equal-mass, non-spinning binaries, we have found that
the results obtained with the most recent, 33-orbit numerical 
waveform are consistent with previous results obtained using the
older, shorter 15-orbit simulation. The errors between hybrids using
these two NR waveforms are small enough for parameter estimation for
sources with $\rho_\text{eff}<100$. In addition, when compared to
PN, these two NR waveforms yield similar results. The
results in Fig.~\ref{fig:T1vsT2vsT3vsT4_loglog_q1} show that the
results using the 15-orbit waveform can be reproduced and extended.

Expanding the parameter space to unequal-mass binaries, we have
found that PN errors grow with increased mass ratio as observed
by~\cite{Boyle:2011dy,Hannam:2010,OhmeEtAl:2011,Santamaria:2010yb}. Phase
differences between PN and NR vary strongly with mass ratio and PN
approximant. Therefore, no single PN waveform is appropriate for all of parameter
space. For example, despite the fact that Taylor T4 matches remarkably
well with NR in the case of equal-mass, non-spinning binaries, this is
no longer true as the mass ratio between the black holes in the binary
system increases. It is only a coincidence that TaylorT4 and NR are so
similar in this very unique configuration. In fact, for $q = 6$,
TaylorT1 and TaylorT2 agree very well with NR, whereas TaylorT4 does
not, cf. Fig.~\ref{fig:AccumDiff}.  

When evaluating $\Q$ for hybrids which use different PN approximants,
it becomes clear that higher mass ratio binaries will require
increasingly more NR orbits to reach similar accuracy. This becomes
problematic for high mass ratio binaries, since they are more
computationally expensive because it requires more numerical
resolution to resolve the smaller black hole.  We also note the
  recent analysis~\cite{Lovelace:2011nu} of an equal-mass aligned spin
  BBH simulation with very large spins.  Ref.~\cite{Lovelace:2011nu}
  found that the simulation covering $\sim 25$ orbits was of
  insufficient length to reliably hybridize with PN.

Knowledge of higher-order PN waveforms will significantly improve the
quality of PN+NR hybrid waveforms.  As demonstrated in
Figs.~\ref{fig:dtbiasVSomegaT4} and~\ref{fig:dtbiasVSomegaT4_q6}, any
further additional PN order should decrease errors in our measure 
$\Q$ by roughly a factor of $\sim 2$, for the same length of the NR
waveform.  Because mismatches are proportional to the square
of $\Q$ this will reduce mismatches by a factor $\sim 4$.  For a
specified accuracy, the knowledge of a higher-order PN expansion would allow one to shorten
the length of NR simulations. For $\rho_{\rm eff}=10$,
Fig.~\ref{fig:dtbiasVSomegaT4} indicates that the matching frequency
could be raised from $M\omega_m=0.027$ to $M\omega_m=0.038$, thus
approximately reducing the temporal length of the NR simulation by a
factor of $\sim 2$.  This would be a substantial saving for future NR
simulations.  We emphasize that these estimates depend on our assumption
of the approximate magnitude of the unknown PN coefficients, as discussed
in the context of Fig.~\ref{fig:A_PN_vs_PNorder}.

This analysis has also provided us with an important cautionary tale:
one cannot ignore outliers when comparing many different PN
approximants to each other. This is illustrated in the 3PN case in
Fig.~\ref{fig:T1vsT2vsT3vsT4_30_30_q1}, where the error between Taylor
T2, T3, and T4 hybrids is very small, but when compared to Taylor 
T1 hybrids, the
error is much higher. Thus, it is very important, when doing this type
of analysis, to consider as many different PN approximants as possible.

This work could be extended in a few ways. First of all, it would be
interesting to further extend the parameter space to spinning
binaries, and also to the most general case of precessing
binaries. Some work has been done to create hybrids for precessing
binaries in~\cite{CampanelliEtal2009, Schmidt2010, OShaughnessy2011,
  Boyle:2011gg}, but error estimates on this type of hybrid are still
a ways off.  It would also be useful to extend the error
  analysis of GW modes to modes different from just the (2,2) mode,
  because these other modes become increasingly important with higher
  mass ratio (e.g.,~\cite{Buchman:2012dw}) and with precession of the
  orbital plane (e.g.,~\cite{Buonanno:2002fy, PekowskyEtAl:2012}).

It would also be useful to refine our error limit in
Eq.~\ref{eq:dh-over-h-first} to incorporate the effects of a network
of detectors in a more effective way. It might be worthwhile to have
an error limit which depends on the total mass of the binary as
in~\cite{Boyle:2011dy}, or to consider some better way of finding an
upper bound on $\Q$. An important consideration is the fact that our
error criterion is sufficient but not necessary to the suitability of
these hybrid waveforms.  For instance, investigations into detection
efficiency of hybrid waveforms~\cite{Ohme:2011rm,OhmeEtAl:2011}
indicate that hybrid waveforms can be perfectly usable, even when
failing the indistinguishability test by a wide margin.

\begin{acknowledgments}
  We would like to thank Mark Hannam, Sascha Husa, and
  Ulrich Sperhake for useful discussions, and thank Riccardo Sturani,
  Stefano Foffa, Luc Blanchet, Chad Galley, Alessandra Buonanno, and
  Samaya Nissanke for insights into the difficulties in computing
  higher order post-Newtonian expansions.  The numerical waveforms
  used in this work were computed with the SpEC
  code~\cite{SpECwebsite}.  We gratefully acknowledge support from the
  NSERC of Canada, from the Canada Research Chairs Program, from the
  Canadian Institute for Advanced Research, and from the Sherman
  Fairchild Foundation; from NSF grants PHY-0969111 and PHY-1005426 at
  Cornell, and from NSF grants PHY-1068881 and PHY-1005655 at Caltech.
  Computations were performed on the Zwicky cluster at Caltech,
    which is supported by the Sherman Fairchild Foundation and by NSF
    award PHY-0960291; on the NSF XSEDE network under grant
  TG-PHY990007N; and on the GPC supercomputer at the SciNet HPC
  Consortium~\cite{scinet}. SciNet is funded by: the Canada Foundation
  for Innovation under the auspices of Compute Canada; the Government
  of Ontario; Ontario Research Fund--Research Excellence; and the
  University of Toronto.
\end{acknowledgments}


\bibliography{References/References}

\end{document}